\documentclass[sigconf]{acmart}
\usepackage{graphicx} 
\usepackage[english]{babel}
\usepackage{subcaption}
\usepackage{siunitx}
\usepackage{xcolor}
\usepackage{listings}
\usepackage{multirow}
\usepackage{colortbl}
\usepackage{makecell}
\AtBeginDocument{%
  \providecommand\BibTeX{{%
    \normalfont B\kern-0.5em{\scshape i\kern-0.25em b}\kern-0.8em\TeX}}}

\copyrightyear{2024} 
\acmYear{2024} 
\setcopyright{acmlicensed}
\acmConference[ICPC '24]{32nd IEEE/ACM International Conference on Program Comprehension}{April 15--16, 2024}{Lisbon, Portugal}
\acmBooktitle{32nd IEEE/ACM International Conference on Program Comprehension (ICPC '24), April 15--16, 2024, Lisbon, Portugal}
\acmDOI{10.1145/3643916.3644395}
\acmISBN{979-8-4007-0586-1/24/04}

\begin{document}
\title{Exploring the Impact of Source Code Linearity on the Programmers' Comprehension of API Code Examples}
\author{Seham Alharbi}
\orcid{0000-0002-1436-7992}
\authornote{Seham Alharbi is also affiliated with the College of Computer, Qassim University, Buraydah, Saudi Arabia.}
\email{saaa528@york.ac.uk}
\affiliation{%
  \institution{University of York}
  \city{York}
  \country{United Kingdom}
}
\author{Dimitris Kolovos}
\orcid{0000-0002-1724-6563}
\email{dimitris.kolovos@york.ac.uk}
\affiliation{%
  \institution{University of York}
  \city{York}
  \country{United Kingdom}
}
\renewcommand{\shortauthors}{Alharbi and Kolovos}
\newcommand{\dk}[1]{{\color{teal} [DK: #1]}}
\begin{abstract}
\textbf{Context:} Application Programming Interface (API) code examples are an essential knowledge resource for learning APIs. However, a few user studies have explored how the structural characteristics of the source code in code examples impact their comprehensibility and reusability. \\
\textbf{Objectives:} We investigated whether the (a) linearity and (b) length of the source code in API code examples affect users' performance in terms of correctness and time spent. We also collected subjective ratings.\\
\textbf{Methods:} We conducted an online controlled code comprehension experiment with 61 Java developers. As a case study, we used the API code examples from the Joda-Time Java library. We had participants perform code comprehension and reuse tasks on variants of the example with different lengths and degrees of linearity. \\
\textbf{Findings:} Participants demonstrated faster reaction times when exposed to linear code examples. However, no substantial differences in correctness or subjective ratings were observed. \\
\textbf{Implications:} Our findings suggest that the linear presentation of a source code may enhance initial example understanding and reusability. This, in turn, may provide API developers with some insights into the effective structuring of their API code examples. However, we highlight the need for further investigation.
\end{abstract}
\maketitle
\section{Introduction}
\label{section:1}
Working through Application Programming Interface (API) code examples has been proven to be the most preferred learning strategy for both beginner and experienced API users \cite{meng2018application}. Surprisingly, little is known about how the different source code structures in these examples affect their comprehensibility and reusability. Furthermore, existing work (discussed in Section \ref{section:5}) appears to be focused on examining the impact of several source code characteristics on comprehension only in the context of generic software. Therefore, to fill this gap, the study presented in this paper focuses specifically on API code examples. Moreover, unlike existing studies, we examine different source code constructs that illustrate the same API usage and functionality, narrowing the evaluation focus to the constructs' impact on comprehension. We also assess an additional concept, which is the impact of the examined source code structures on the reusability of API code examples.
\par In this study, we are particularly interested in exploring the impact of two source code aspects: the degree of linearity and length. Linear source code refers to code that can be read primarily in a sequential order without interference from interdependent methods or classes. Considering the absence of jumps between method definitions in such a code, we hypothesise it may be easier to comprehend. In addition, linear API code examples may be easier to reuse and adapt into one's codebase, as they typically contain a single self-contained method that can be copied and edited, as opposed to non-linear code examples that involve multiple methods.
\par Through our study, we aim to help API developers understand how to structure their API code examples more effectively, thereby enhancing the examples' comprehensibility and reusability. This, in turn, would promote the learnability of their APIs.
\par This work is a part of larger research \cite{Seham2022} in which we develop tools and techniques to enhance the maintainability and comprehensibility of API code examples. All data collected or used in this study is available in our replication package.\footnote{\label{RP}Replication package: \url{https://figshare.com/s/52e11ece2f39bac64bcb}}
\section{Methodology}
\label{section:2}
\subsection{Research Questions}
\label{section:2.1}
We intended to answer the following research questions:
\begin{description}
\item[\textbf{RQ1:}] In terms of correctness and time spent, how does the linearity of an API code example impact a programmer's performance in tasks that require code comprehension?
  \item[\textbf{RQ2:}] What effects does the length of a linear API code example\footnote{\label{f2}Please refer to our (\textcolor{blue}{\hyperref[RP]{replication package}} \(\rightarrow\) examples) for some sample code illustrating linear and non-linear API code examples.} have on its comprehensibility and reusability?
  \item[\textbf{RQ3:}] Does the degree of linearity in a non-linear API example affect its comprehensibility and reusability?
\end{description}
\subsection{Study Design}
\label{section:2.2}
This study was conducted online to allow access to a large and more diverse pool of participants. We utilised Gorilla \cite{gorilla2023}, a widely used online experiment builder, which provided all the features we needed in our study (randomisation, counterbalancing presentation of formatted source code, accurate reaction times and integration with participant recruitment platforms). We recruited Java developers from Prolific\footnote{\url{https://www.prolific.com}} -- a participant recruitment platform for online research -- and assigned them to two main groups: linear vs non-linear (between-subjects). Each group consisted of two sub-groups that correspond to the treatment categories shown in Table \ref{t2}. Participant assignment to groups was fully randomised and balanced, with a 1:1 ratio. Each participant completed two code comprehension and reuse tasks from the same treatment category (highlighted in the same colour in Table \ref{t2}). The order in which each participant received tasks was also randomised to eliminate any potential order effects. Ethical approval was obtained before the study was conducted.
\subsection{Independent Variables}
\label{section:2.3}
We considered a single independent variable: the source code structure of API code examples. Two primary source code factors were systematically varied: code linearity, which is manipulated using the source code linearity metric (\textit{i}) proposed by Peitek et al. \cite{Norman2020what}; and code length, which is varied by adjusting the number of lines of code (LOC). As shown in Table \ref{t2}, we combined these two factors and generated four treatment categories: \setlength{\fboxsep}{0pt}\colorbox{yellow!10}{linear-short,}  \colorbox{teal!10}{linear-long,} and non-linear with varying levels of linearity \((i)\), ranging from \colorbox{red!10}{\((10.00 < i \leq 15.00)\)} to \colorbox{orange!10}{\((15.00 < i \leq 20.00)\)}. These selected values reflect a diverse spectrum of code linearity.
\subsection{Dependent Variables}
\label{section:2.4}
We measured three dependent variables: reaction time, correctness and subjective rating.
\par \textit{Time Duration Marking.} For the comprehension phase, reaction time was defined as the amount of time that elapsed between a participant’s initial view of an API code example and submission of their overall comprehension rating Similarly, for the code-reuse phase, reaction time was defined as the time between a participant's first view of the required code-reuse task and the submission of their solution code.
\par \textit{Judging Correctness.} We ensured marking consistency by defining a set of correctness categories and criteria\footnote{Detailed criteria are available in our \textcolor{blue}{\hyperref[RP]{replication package}}.}: correct (A), almost correct (B), partially correct (C), incorrect (D) and absent (F).
\begin{table*}[!ht]
    \setlength{\tabcolsep}{5pt}
    \footnotesize
    \caption{API code examples and their variants, metric values and study results. Variants sharing the same colour belong to the same treatment category. The imbalance in the number of responses (N) between two variants of the same example is due to our exclusion of responses with inaccurately reported break times.}
    \label{t2}
    \begin{tabular}{llllll|llllll}
    \toprule
    \makecell{\textbf{API} \\ \textbf{Code Example}} & \textbf{Variant} & \multicolumn{4}{c}{\textbf{Metrics}} & \textbf{N} & \textbf{Correctness} & \multicolumn{2}{c}{\textbf{Comprehension}} & \multicolumn{2}{c}{\textbf{Reuse}} \\
    & & LOC & Complexity & Linearity \textit{(i)} & \makecell{\# Method \\ Calls} &  &  & \makecell{Median \\ Reaction Time} & p-value$^{\mathrm{*}}$ & \makecell{Median \\ Reaction Time} & p-value$^{\mathrm{*}}$ \\
    \midrule
    Date Example &  \cellcolor{yellow!10}Linear & \cellcolor{yellow!10}31 & \cellcolor{yellow!10}7 & \cellcolor{yellow!10}0.00 & \cellcolor{yellow!10}0 & 17 & 15 (88\%) & 44s & \multirow{2}{*}{\textbf{0.01}} & 6m 11s & \multirow{2}{*}{0.36} \\
    & \cellcolor{red!10}Non-linear & \cellcolor{red!10}14 & \cellcolor{red!10}1 & \cellcolor{red!10}12.95 & \cellcolor{red!10}5 & 13 & 8 (62\%) & 1m 38s &  &  7m 14s & \\
    Chronology Example & \cellcolor{yellow!10}Linear & \cellcolor{yellow!10}25 & \cellcolor{yellow!10}4 & \cellcolor{yellow!10}0.00 & \cellcolor{yellow!10}0 & 18 & 10 (56\%) & 43s & \multirow{2}{*}{\textbf{0.03}} & 3m 37s & \multirow{2}{*}{\textbf{0.10}} \\
    & \cellcolor{red!10}Non-linear & \cellcolor{red!10}11 & \cellcolor{red!10}1 & \cellcolor{red!10}11.41 & \cellcolor{red!10}4 & 14 & 13 (93\%) & 1m 49s & & 6m 15s & \\
    Duration Example & \cellcolor{teal!10}Linear & \cellcolor{teal!10}35 & \cellcolor{teal!10}7 & \cellcolor{teal!10}0.00 & \cellcolor{teal!10}0 & 16 & 9 (56\%) & 3m 31s & \multirow{2}{*}{0.60} & 16m 18s & \multirow{2}{*}{0.40} \\
    & \cellcolor{orange!10}Non-linear & \cellcolor{orange!10}21 & \cellcolor{orange!10}2 & \cellcolor{orange!10}19.47 & \cellcolor{orange!10}6 & 12 & 5 (42\%) & 2m 17s &  & 24m 59s & \\
    Interval Example & \cellcolor{teal!10}Linear & \cellcolor{teal!10}46 & \cellcolor{teal!10}7 & \cellcolor{teal!10}0.00 & \cellcolor{teal!10}0 & 14 & 10 (71\%) & 1m 14s & \multirow{2}{*}{0.33} & 9m 21s & \multirow{2}{*}{\textbf{0.06}} \\
    & \cellcolor{orange!10}Non-linear & \cellcolor{orange!10}33 & \cellcolor{orange!10}5 & \cellcolor{orange!10}19.43 & \cellcolor{orange!10}6 & 12 & 9 (75\%) & 2m 49s &  & 12m 56s & \\
    \midrule
    \textbf{Overall} & \multicolumn{2}{l}{\textbf{Linear}} & & &  & 65 & 44 (68\%) & 58s &  & 7m 17s &  \\
    & \multicolumn{2}{l}{\textbf{Non-linear}} & & &  & 51 & 35 (69\%) & 1m 50s &  & 10m 13s &  \\
    \bottomrule
    {$^{\mathrm{*}}$ Mann--Whitney U test}
\end{tabular}
\end{table*}
\subsection{Participants}
\label{section:2.5}
\textit{Pilot.} To validate the study design, we conducted a pilot with four participants (average age 28.5 ± 9.5; average years of Java programming 3.2 ± 1.5). Based on the results, we reduced the number of tasks assigned to each participant from four to two to minimise the experiment’s overall duration. We also changed the online IDE\footnote{\url{https://replit.com}} used due to its slow execution time and enhanced the wording of the code optimisation question. The data collected in the pilot study was not used in the final analysis.
\par \textit{Pre-screening survey.} In addition to the pre-screeners provided by Prolific, we created a separate programming knowledge survey to assess participants' programming knowledge before they participated in the study. This was essential since recent research revealed that, while recruitment platforms, such as Prolific, greatly mitigate self-selection bias and some security issues, their pre-screeners may not always be reliable \cite{Danilova2021, Russo2022, Wagner2022}. In this survey, we used the basic knowledge questions and time limit recommended by Danilova et al. \cite{Danilova2021}. Participants who correctly answered all the questions within the time limit were manually invited to participate in the study.
\par \textit{Study.} We recruited 61 Java developers from 14 countries with varying levels of programming experience. Only 19 (31\%) participants stated that they previously used the Joda-Time Java library. Among them, only eight (42\%) said that they used it more than once, and none of them reported regular usage. The participants' prior programming experience was assessed using a validated questionnaire that is based on self-estimates \cite{janet2012, janet2014}. Each participant was compensated £10 for their time and effort. Additional information about the participant demographics is shown in Table \ref{t1}.
\begin{table}[!ht]
    \setlength{\tabcolsep}{20pt}
    \footnotesize
    \caption{Participant demographics.}
    \label{t1}
    \begin{tabular}{lcc}
    \toprule
    \textbf{Category} & \textbf{n=61} \\
    \midrule
    \textbf{Student} & 43 (70\%) \\
    \textbf{Professional Developer} &   18 (30\%) \\
    \textbf{Programming Experience} (in Years) &  5.9 ± 3.3 \\
    \textbf{Java Programming Experience} (in Years) &  3.2 ± 2.2 \\
    \textbf{Familiarity with Joda-Time} &   19 (31\%) \\
    \textbf{Male} &   54 (89\%) \\
    \textbf{Female} &   7 (11\%) \\
    \textbf{Age} (in Years) &  25.4 ± 6.1  \\
    \bottomrule
\end{tabular}
\end{table}
\subsection{Material} 
\label{section:2.6}
\textit{API code examples.} We chose the Joda-Time\footnote{\label{joda-time}\url{https://www.joda.org/joda-time/}} Java library because it addresses a well-known concept (i.e. date and time handling). Joda-Time met our selection criteria of: 1) not requiring prior domain knowledge that would pose an unnecessary challenge to participants; 2) being well-documented and 3) not being too popular so that an average Java developer would not necessarily be familiar with it. Furthermore, we intended to utilise the code examples available on the Joda-Time documentation page.\hyperref[joda-time]{\textsuperscript{\ref*{joda-time}}} However, we found that these examples were not complex enough. Thus, we decided to create our own examples.
\par As shown in Table \ref{t2}, we developed four API code examples, each of which demonstrated a distinct usage of Joda-Time. This variation of examples was important to minimise the risk of any potential learning effect arising from within the examples. We then created linear and non-linear versions of each example. We strove to make these examples look as natural as possible by: 1) properly documenting them and 2) letting them depict real-world scenarios such as data manipulation or meeting scheduling. For the non-linear versions of the examples, we refactored the linear version by extracting some functionalities into a set of utility methods and replacing the extracted code with method calls using the extract method refactoring technique. Examples within the same treatment category were relatively comparable in terms of length, complexity, degree of linearity and the number of utility method calls.
\par \textit{Tasks.} Oftentimes, when API users turn to code examples to learn a new API, they typically have a specific problem in mind. They are hoping that the code in the example they are reviewing will be reusable. If this is possible, they copy and paste the example, then modify its source code by adding or deleting statements to match their needs \cite{meng2018application, Gao2020}. In our study, we wanted to simulate this behaviour. Therefore, each code-reuse task had two parts: 1) code modification, in which participants were required to make changes to address a specific problem; and 2) code optimisation, in which we asked them to remove any unnecessary code that did not directly contribute to their task solution. The tasks were generally easy and designed to be solved with a few edits. Each API code example had a unique task that remained the same for both versions of the example.
\subsection{Experiment Procedure}
\label{section:2.7}
After obtaining their consent, we asked the participants to complete a demographics questionnaire. Subsequently, each of them was randomly assigned two code examples from the same treatment category. This means that each participant completed two distinct code-reuse tasks.
\par We designed each task to be completed in four sequential parts. The first was the comprehension part, in which participants were asked to review the example and rate their own understanding. The next part pertained to instructions, in which participants were given a link to an online IDE\footnote{\url{https://www.jdoodle.com}} that contained a Java project of the example, with Joda-Time imported and ready to use. Participants were also instructed on how to download the example if they preferred using their own IDE. The third part was the code-reuse task, in which participants answered a two-part question (as explained in Section \ref{section:2.6}) and pasted their solution code in a given text box. The final part involved post-task questions, in which participants were asked to rate how difficult it was to reuse the code, whether they employed the provided online IDE or their own, and report any break time (if any was taken).
\par We only measured the time spent on two of the four parts: 1) comprehension time and 3) reuse time. The rationale for separating the comprehension and reuse of the same code example was to reduce participants' use of the `as-needed' program comprehension strategy \cite{littman1987, anneliese1998, Dror2021}.
\subsection{Data Analysis}
\label{section:2.8}
We manually analysed the correctness of responses for each task. First, we converted the categories mentioned in Section \ref{section:2.4} to numerical values (correct (A) = 100\%, almost correct (B) = 70\%, partially correct (C) = 40\%; both incorrect (D) and absent (F) = 0\%). We applied the same scale to both parts of the code-reuse task (i.e. code modification and code optimisation). However, when calculating the overall task score, we assigned more weight (90\%) to the first part of the task, as it required greater effort than the second part, which accounted for only 10\% of the total weight. Responses with an overall score of 60\% or higher were considered correct. This overall correctness threshold ensures that participants achieve at least 70\% in the first part of the task.
\par When analysing reaction times automatically captured by Gorilla \cite{gorilla2023}, we only considered correct responses. We used the Shapiro-Wilk test \cite{shapiro1965} to assess the normality of reaction times and correctness as well as Levene's test \cite{Levene1960} to evaluate variance homogeneity. The findings indicated non-normality and unequal variances for both correctness and reaction times. Therefore, to test for statistically significant differences, we used a non-parametric test, the Mann--Whitney U test (Wilcoxon rank-sum test), with a significance level of \(\alpha = 0.05\).
\section{Results and Discussion}
\label{section:3}
\begin{figure}[!ht]
\subfloat[Comprehension time]{\includegraphics[width=0.5\columnwidth]{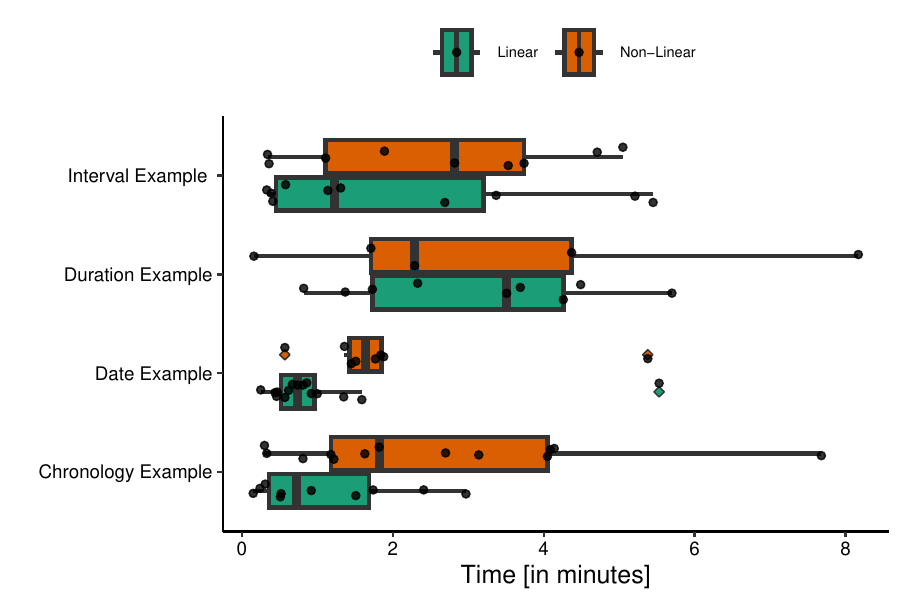}}
\subfloat[Reuse time]{\includegraphics[width=0.5\columnwidth]{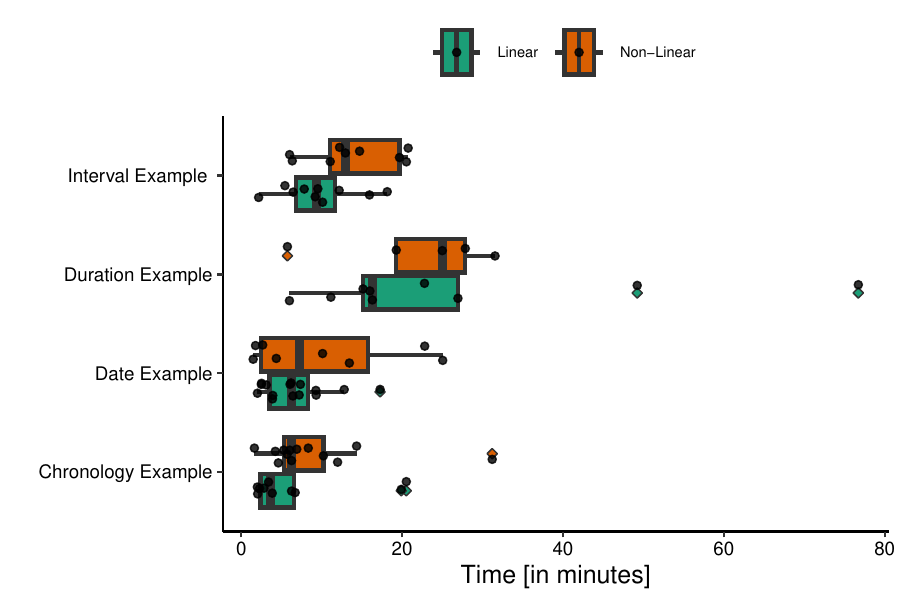}}
\caption{The time spent on (a) comprehending, and (b) reusing the API code examples used in the study. Each boxplot represents the responses for one version (linear or non-linear) of a single example.}
\label{fig:1}
\end{figure}

\begin{figure}[!ht]
\subfloat[Comprehension Ratings]{\includegraphics[width=0.5\columnwidth]{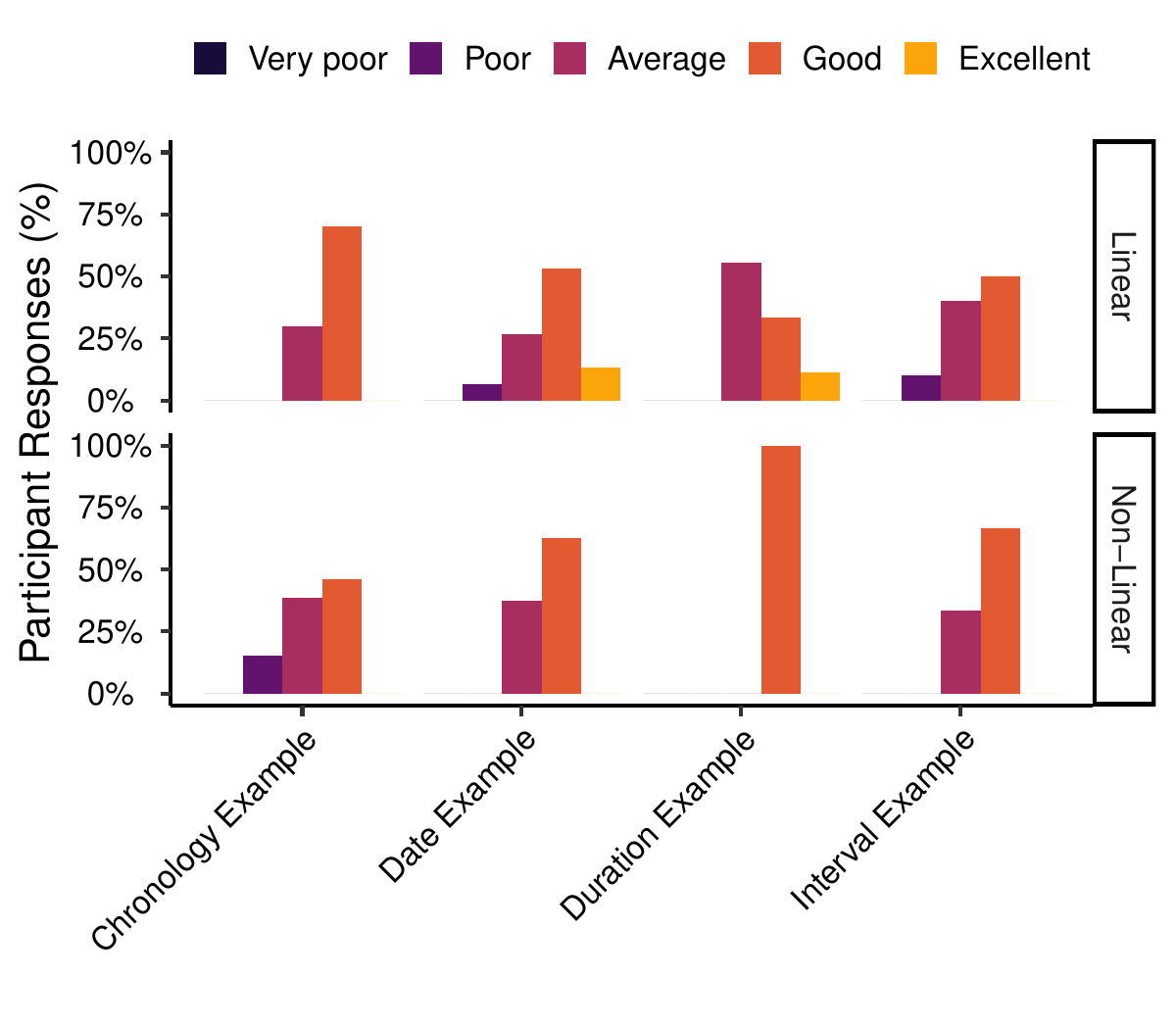}}
\subfloat[Reusability Difficulty Ratings]{\includegraphics[width=0.5\columnwidth]{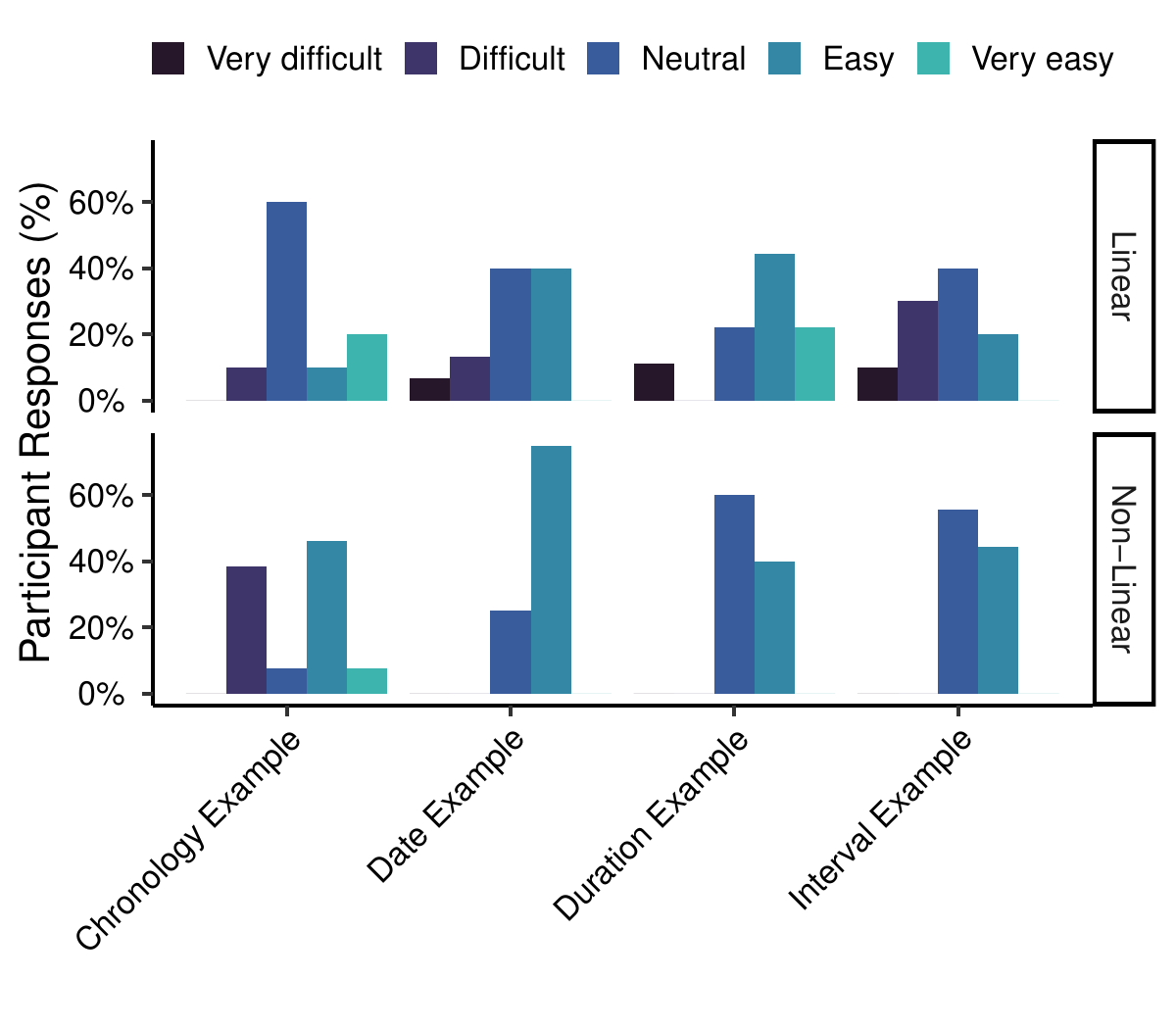}}
\caption{Participants' subjective ratings of API code example comprehension (a) and reusability difficulty (b).}
\label{fig:2}
\end{figure}

As shown in Table \ref{t2} and Figure \ref{fig:1}, participants generally spent less time comprehending and reusing linear code examples (both in mean and median reaction times). This observation suggests that the source code linearity in an API code example may affect a programmer's performance. This influence has a greater impact on comprehension and is statistically significant when the linear API code example is also short (e.g. date and chronology examples). Moreover, in terms of reusability, there appears to be a trend towards significance in two of the API code examples (chronology and interval examples), as reflected by their moderate p-values of \num{0.10} and \num{0.06}, respectively.  However, the impact on correctness (Mann--Whitney U test, \(W = 1690, p = 0.428\)) and subjective rating (as shown in Figure \ref{fig:2}) was not substantial \textbf{(RQ1)}.
\par The Mann--Whitney U test revealed a significant difference in both comprehension (\(W = 125, p = 0.004\)) and  reusability (\(W = 99, p = 0.000\)) between the groups that received linear-short and linear-long API code examples \textbf{(RQ2)}. Similarly, participants spent less time reusing the non-linear code examples when the linearity value (\textit{i}) was lower (\(i < 15.00\), Mann--Whitney U test: \(W = 68, p = 0.003\)). Notably, unlike the comparison in RQ1, this comparison is based on code examples illustrating different API usage; thus, the significant differences in participants' performance could be due to variations in the implemented API functionality and required tasks \textbf{(RQ3)}.
\section{Threats to Validity}
\label{section:4}
\textit{Construct Validity.} To mitigate construct validity threats, we used specific metrics to manipulate our independent variables. We created code examples and tasks that represented real-world scenarios and ensured a consistent correctness evaluation. However, since the experiment was conducted online, we had limited control over participants' activities. To address this, we asked them to self-report break times, which were subtracted from the time spent on solving the task, and disclose whether they used their own IDEs. While this approach provided insights into participants' behaviour, it was not entirely conclusive.
\par \textit{Internal Validity.} We reduced internal validity threats by randomly assigning participants to treatment groups and randomising the order in which they viewed tasks. We also administered a validated programming experience questionnaire and pre-screened participants for their programming knowledge.
\par \textit{External Validity.} One potential threat to external validity was the study’s limited scope. It focused solely on the API of one Java library and included only a few code examples. Also, 70\% of the participants were students, which limited the generalisability of our findings. The study's virtual setting and the use of an online IDE, which lacks features such as auto-completion and error checking, may not fully reflect the conditions of a traditional coding environment.
\section{Related Work}
\label{section:5}
A large number of studies investigated the impact of various source code characteristics on programmers' code comprehension. These characteristics include the use of intermediate variables \cite{cates2021}, certain syntactic structures such as \texttt{ifs} and \texttt{for} loops \cite{Shulamyt2019}, different identifier names \cite{Johannes2019, Eran2017} as well as naming conventions \cite{Bonita2010, Dave2013}. Moreover, some studies explored the effects of more global factors such as the order of methods \cite{Yorai2016}, code regularity \cite{Ahmad2014}, and the linearity of source code and reading order \cite{Norman2020what}.
\section{Conclusion and Future Work}
\label{section:6}
In this paper, we investigated the impact of source code linearity and length on the comprehensibility and reusability of API code examples. We chose code examples from the Joda-Time Java library and manipulated their structure. Furthermore, we recruited 61 Java developers, assigned each one of them two code examples, and asked them to complete code-reuse tasks. This study found that participants demonstrated relatively faster reaction times when working with linear API examples.
\par For future work, we intend to expand this study by incorporating a broader range of APIs from diverse domains. This will involve utilising a larger set of code examples with varying levels of linearity and increasing the number of participants. Also, to better capture participants' activities, we plan to conduct this experiment in a laboratory setting. Additionally, we are interested in determining whether the activities of participants, as they work with code examples of different linearity, still align with the activities reported in existing studies on the COIL\footnote{Collection and Organization of Information for Learning.} model \cite{Gao2020, soren2023}.
\vspace{1cm}
\bibliographystyle{ACM-Reference-Format}
\bibliography{references}
\appendix
\end{document}